\documentclass[acmsmall]{acmart}
\AtBeginDocument{%
  }

\setcopyright{cc}
\setcctype{by}
\acmDOI{10.1145/3808207}
\acmYear{2026}
\acmJournal{PACMSE}
\acmVolume{3}
\acmNumber{FSE}
\acmArticle{FSE200}
\acmMonth{7}
\acmSubmissionID{fse26mainb-p2722-p}
\received{2025-09-12}
\received[accepted]{2026-03-24}

\usepackage{amsmath}
\usepackage{booktabs}
\usepackage{multirow}
\usepackage{makecell}
\usepackage{threeparttable}
\usepackage{enumitem}
\usepackage{balance}

\usepackage{xcolor}
\usepackage{graphicx}
\usepackage{pifont}
\usepackage{wasysym}

\usepackage{listings}
\usepackage{lstlinebgrd}
\usepackage[linesnumbered,ruled,vlined]{algorithm2e}

\usepackage{caption}
\usepackage{subcaption}
\usepackage{tcolorbox}
\usepackage{fancybox}

\usepackage{mathpartir}
\usepackage{xspace}
\usepackage{url}

\usepackage{minted}

\usepackage{hyperref}
\usepackage{siunitx}

\newcommand{\eg}{e.g.}

\newcommand{\ccSysName}[0]{TEERepair\xspace}
\newcommand{\ccBenchName}[0]{PartitioningE-Bench\xspace}
\newcommand{\ccTitle}[0]{Automated Repair of TEE Partitioning Issues via DSL-Guided and LLM-Assisted Patching}

\newcommand{\blackding}[1]{\ding{\numexpr181+#1\relax}}

\definecolor{codebg}{rgb}{0.99,0.99,0.99}
\newminted[CSourceCode]{c}{
    breaklines,
    bgcolor=codebg,
    labelposition=topline,
    frame=single,
    numbersep=2pt,
    fontsize=\scriptsize,
    xleftmargin=5pt,
    framerule=0.5pt,
    linenos,
    escapeinside=||
}

\newminted[TextSourceCode]{text}{
    breaklines,
    labelposition=topline,
    frame=single,
    numbersep=2pt,
    fontsize=\scriptsize,
    xleftmargin=5pt,
    framerule=0.5pt,
    linenos,
    escapeinside=||
}

\newminted[MarkdownSourceCode]{markdown}{
    breaklines,
    labelposition=topline,
    frame=single,
    numbersep=2pt,
    fontsize=\scriptsize,
    xleftmargin=5pt,
    framerule=0.5pt,
    linenos,
    escapeinside=||
}

\definecolor{R1color}{RGB}{255,100,25}
\definecolor{R2color}{RGB}{50,100,255}
\definecolor{R3color}{RGB}{100,100,100}
\definecolor{R4color}{RGB}{255,255,100}

\lstdefinelanguage{DSL}{
  morekeywords={RULE,MATCH,WHEN,REWRITE,BIND,POST,ASSERT,IF,
    COPY,MALLOC,ENC,READ,WRITE,HASH,ARRAY,SHALLOW,MUTATE,RAW,
    len,size,equal,PASS,return,SNPRINT,MULMALLOC,MULENC},
  sensitive=true,
  morecomment=[l]{//},
  morestring=[b]",
}

\lstdefinelanguage{TEE}[]{C}{
  breaklines,
  morekeywords={TEE_MemMove,TEE_Malloc,TEE_MemCompare,TEE_ERROR_BAD_PARAMETERS,TEE_SUCCESS,enc,snprintf,hash,read,write},
  sensitive=true,
}

\newcommand\ans[1]{
\vspace{5pt}
\noindent
\doublebox{
    \begin{minipage}{0.96\textwidth}
      #1
    \end{minipage}
  }
}
\begin{document}

\title{\ccTitle}


\author{Chengyan Ma}
\orcid{0000-0001-9256-6930}
\affiliation{%
  \institution{Singapore Management University}
  \city{Singapore}
  \country{Singapore}
}
\email{chengyanma@smu.edu.sg}

\author{Jieke Shi}
\orcid{0000-0002-0799-5018}
\affiliation{%
  \institution{Singapore Management University}
  \city{Singapore}
  \country{Singapore}
}
\authornote{Corresponding author.}
\email{jiekeshi@smu.edu.sg}

\author{Ruidong Han}
\orcid{0000-0001-6859-6005}
\affiliation{%
  \institution{Singapore Management University}
  \city{Singapore}
  \country{Singapore}
}
\email{rdhan@smu.edu.sg}

\author{Ye Liu}
\orcid{0000-0001-6709-3721}
\affiliation{%
  \institution{Singapore Management University}
  \city{Singapore}
  \country{Singapore}
}
\email{yeliu@smu.edu.sg}

\author{Feng Li}
\orcid{0000-0002-8294-7606}
\affiliation{%
  \institution{Singapore Management University}
  \city{Singapore}
  \country{Singapore}
}
\email{fengli@smu.edu.sg}

\author{Yuqing Niu}
\orcid{0009-0003-6794-4970}
\affiliation{%
  \institution{Singapore Management University}
  \city{Singapore}
  \country{Singapore}
}
\email{yuqingniu@smu.edu.sg}

\author{David Lo}
\orcid{0000-0002-4367-7201}
\affiliation{%
  \institution{Singapore Management University}
  \city{Singapore}
  \country{Singapore}
}
\email{davidlo@smu.edu.sg}

\renewcommand{\shortauthors}{Chengyan Ma, Jieke Shi, Ruidong Han, Ye Liu, Feng Li, Yuqing Niu, and David Lo}

\begin{abstract}
Trusted Execution Environments (TEEs) provide hardware-based isolation to protect sensitive data and computations from potentially compromised operating systems (OS). However, TEE applications inevitably interact with the untrusted OS through SDK interfaces, and improper partitioning can introduce severe vulnerabilities such as data leakage and code injection. While prior work has proposed static analysis tools to detect such issues, automated repair remains largely unexplored. This problem is particularly challenging due to three TEE-specific factors: the lack of standardized secure development guidelines, the difficulty of extracting semantic information from low-level C code, and the absence of mature testing and validation methods.
In this work, we present \ccSysName, a framework for automatically repairing bad partitioning issues in TEE applications. Our approach tackles the above challenges by introducing a domain-specific language (DSL) to encode repair rules that express and capture common TEE security patterns, which are instantiated as patch templates with placeholders for context-specific variables. We then leverage large language models (LLMs) to reason about code semantics and synthesize context-aware patches, and further generate test clients to validate the repairs. We evaluate \ccSysName on the TEE Partitioning Errors Benchmark (\ccBenchName), achieving a significantly higher repair success rate of 87.6\% compared to baselines. Furthermore, applying \ccSysName to real-world TEE projects, we submitted 5 repair pull requests, 2 of which have been confirmed and merged by project maintainers.
\end{abstract}




\maketitle

\section{Introduction}
\label{sec:intro}
\begin{figure}[!t]
\centering
\begin{minipage}[c]{0.325\textwidth}
\centering
\includegraphics[width=\linewidth]{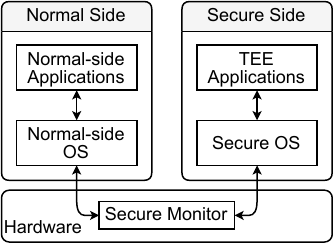}
\end{minipage}
\hspace{0.02\textwidth}
\begin{minipage}[c]{0.635\textwidth}
\centering
\includegraphics[width=\linewidth]{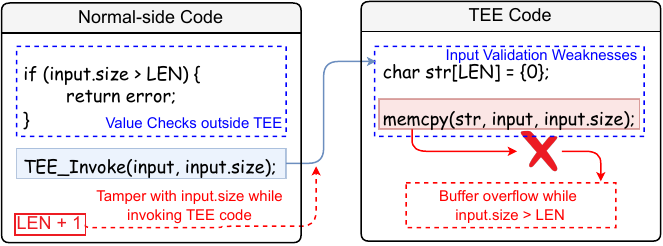}
\end{minipage}
\begin{minipage}[t]{0.325\textwidth}
\centering
\caption{System architecture of TEE.}
\label{fig:tee_arch}
\end{minipage}
\hspace{0.02\textwidth}
\begin{minipage}[t]{0.635\textwidth}
\centering
\caption{An example of input validation weaknesses.}
\label{fig:example_bp}
\end{minipage}
\end{figure}

Trusted Execution Environment (TEE) is an isolated area of memory and CPU that segregates sensitive code and data from the untrusted rest of the system by  enforcing encryption and hardware-based isolation mechanisms (e.g., memory access control and secure CPU instructions)~\cite{7807249,10.1145/2994459.2994460}. Figure~\ref{fig:tee_arch} illustrates the architecture of a typical TEE system such as Intel SGX~\cite{zheng2021survey} or ARM TrustZone~\cite{10.1145/3308755.3308761}, where programs are partitioned into two domains: the normal side, which hosts untrusted applications and the operating system, and the secure side, which contains trusted components that handle sensitive operations such as cryptography~\cite{vella2021rv} or key management~\cite{9024053}. Such partitioning ensures only authorized code can access and manipulate sensitive data within TEE while preventing unauthorized reading or tampering by external code, thereby ensuring the confidentiality and integrity of critical computations~\cite{10.1145/3268935.3268938,10.1007/978-3-030-81645-2_10}. Nowadays, TEEs have been widely adopted in critical domains such as mobile payment~\cite{6726262}, digital rights management~\cite{10.1145/3664476.3664486}, and cloud confidential computing~\cite{10628910}, with a market projected to exceed USD 22.3 billion by 2033~\cite{market}.

Although hardware isolated, TEE applications necessarily expose software interfaces for data exchange between the normal side and TEE~\cite{s20041090}, which introduces potential weaknesses, i.e., if these interfaces are improperly designed or implemented, the isolation and confidentiality guarantees of TEE may be compromised. For example, when TEE code omits essential security mechanisms such as boundary checks, encryption, or integrity verification, attackers from the normal side can manipulate input parameters or intercept sensitive outputs. Figure~\ref{fig:example_bp} illustrates a typical example: while the normal-side code validates the buffer length before invoking TEE, the TEE code performs a memory copy without rechecking the size. An attacker who modifies the size parameter can trigger buffer overflows, leading to sensitive data leakage or corruption of secure execution. Similar risks occur when TEE code outputs unencrypted secrets to the normal side or directly relies on shared memory controlled by untrusted processes. These vulnerabilities are collectively referred to as bad partitioning issues~\cite{ma2025ditingstaticanalyzeridentifying}.

Several research efforts have recognized these threats and developed tools to detect bad partitioning issues in TEE applications. In particular, Ma et al.~\cite{ma2025ditingstaticanalyzeridentifying} conducted a pioneering study that systematically classified bad partitioning into three categories: (1) unencrypted outputs from TEE, (2) insufficient input validation inside TEE, and (3) direct reliance on shared memory controlled by the normal side. They developed a static analyzer to detect such issues and identified hundreds of real-world vulnerabilities across open-source TEE projects, highlighting that insecure partitioning can directly lead to secret leakage, code injection, or compromise of critical services. These findings underscore that bad partitioning is not merely a theoretical concern but a prevalent and severe threat in deployed TEE applications.
Yet, the problem of automatically repairing such vulnerabilities has not been systematically studied due to the following challenges:

\textit{\textbf{Challenge 1:} Lack of standard TEE application development guidelines as reliable ground truths.}
Although TEE SDKs~\cite{sgx-guide, optee-guide} document APIs, they do not prescribe secure partitioning practices. Unlike conventional software security with resources such as OWASP guidelines~\cite{9432179}, TEE development lacks widely accepted standards for partitioning logic, validating inputs across trust boundaries, or managing secrets~\cite{niu2026trustinsecuredemystifyingdevelopers}. Developers often rely on ad hoc designs, resulting in inconsistent code styles and error handling~\cite{niu2026trustinsecuredemystifyingdevelopers}, which prevents automated repair tools from relying on a consistent ground truth.

\textit{\textbf{Challenge 2:} Hard to obtain semantic information in the low-level language.}
Most TEE applications are implemented in C, a low-level language lacking memory-safety guarantees. Repairing vulnerabilities such as buffer overflows requires knowledge of semantic information, such as the intended buffer size, valid index range, or which integer field corresponds to which buffer pointer. Such information is rarely explicit in the code and is often only available through manual auditing~\cite{6903573}. Moreover, TEE-specific programs frequently embed these semantics in implicit ways, such as dynamically allocating memory based on an input parameter or encoding buffer lengths inside struct fields. Without recovering such semantics, repair tools risk generating incorrect patches that either truncate valid data or leave vulnerabilities unfixed.

\textit{\textbf{Challenge 3:} Absence of mature testing approaches for TEE applications.}
TEE applications execute in isolated environments that restrict monitoring and debugging. Intel SGX provides limited enclave debugging, while ARM TrustZone exposes results only through normal-side calls, hindering fine-grained runtime tracing. These heterogeneous mechanisms impede universal test client development~\cite{10179302,DUAN2023192}. Our manual analysis of TEE-related GitHub projects shows most TEE applications lack unit tests or regression suites, limiting test-driven repair applicability~\cite{zhang2024systematicliteraturereviewlarge}. Even when tests exist, the absence of trusted oracles (i.e., reliable reference outputs for comparison) makes it difficult to verify that patches eliminate vulnerabilities while preserving semantics. Additionally, TEE development relies on cross-compilation~\cite{optee}: code is compiled with a specific toolchain on a host machine but must be deployed to TEE-enabled hardware for execution, complicating the repair-compile-test cycle.

In this paper, we propose \ccSysName, a repair framework tailored for TEE applications. The core idea is to craft repair templates capturing secure programming patterns while leaving placeholders for semantic knowledge filled by large language models (LLMs). Specifically, we design a domain-specific language (DSL) that allows developers to express repair rules specifying how vulnerable code patterns should be transformed into secure alternatives. \ccSysName maps detected bad partitioning issues to corresponding DSL rules and instantiates repair templates encoding the secure fix structure with placeholders for context-dependent information such as buffer sizes, variable names, or cryptographic parameters.

These placeholders are resolved by the LLM, which infers appropriate variables, constants, and operators from surrounding code context. The DSL contributes structural repair knowledge (e.g., required security checks), while the LLM supplies contextual semantics (e.g., which buffer to check, which key to use), producing syntactically valid and semantically correct patches. After each repair, we reanalyze the code using a bad partitioning detection tool~\cite{ma2025ditingstaticanalyzeridentifying} to ensure no new issues arise. \ccSysName also automatically synthesizes TEE test clients through static analysis to verify that patches remove vulnerabilities while preserving intended functionality.
    
We apply \ccSysName to the TEE Partitioning Errors Benchmark (\ccBenchName)~\cite{ma2025ditingstaticanalyzeridentifying}, which contains 89 cases of unencrypted data output, input validation weaknesses, and direct usage of shared memory, to assess its effectiveness in repairing bad partitioning issues. 
The results show that \ccSysName can achieve an outstanding repair success rate of 87.6\%. To further validate its practicality, we applied \ccSysName to real-world TEE projects collected from GitHub. \ccSysName successfully generated candidate patches for numerous identified issues, from which we submitted 5 pull requests to the corresponding repositories. At the time of writing, 2 pull requests have already been confirmed and merged by project maintainers. These results highlight the real-world applicability of \ccSysName in assisting developers to secure TEE applications and demonstrate that automated repair can integrate into existing development workflows.

\section{Background and Motivation} \label{sec:bg}
\subsection{Trusted Execution Environment}
A Trusted Execution Environment (TEE) provides a hardware-isolated execution context that safeguards sensitive code and data from potentially malicious software running in the untrusted Operating System (OS), often referred to as the normal side. Its security is grounded in three fundamental properties. First, isolation ensures that code and data inside TEE remain inaccessible to the normal side, even if the operating system or user-level applications are fully compromised. Second, integrity guarantees that the code and state within TEE cannot be modified or tampered with by untrusted components. Third, confidentiality prevents unauthorized disclosure of sensitive assets such as cryptographic keys or proprietary algorithms.

Technically, TEE implementations, such as ARM TrustZone and Intel SGX, combine hardware mechanisms (e.g., memory access control, secure context switching) with minimal trusted software components (e.g., a secure OS or enclave runtime) to establish a small trusted computing base (TCB)~\cite{iTrustee, 10.1145/2666620.2666632, optee}. This design significantly reduces the attack surface compared to a full-fledged operating system. In addition, TEEs support secure boot and remote attestation~\cite{10628729}, which enable remote parties to verify that only authorized code is loaded and executed within TEEs. Collectively, these mechanisms enhance the resilience of TEEs against privileged malware, kernel exploits, and even root-level compromises of the normal-side OS.

\subsection{Bad Partitioning Issues in TEE projects}
\label{sec:definition}
In this section, we provide a precise definition of bad partitioning issues in TEEs and explain why they pose significant threats to TEE applications. Following prior studies~\cite{9925569, 10477533, 10.1145/3319535.3363205}, we assume an adversary with full control over the normal-side execution environment, including its operating system, applications, and memory space. The adversary can arbitrarily tamper with parameters exchanged between the normal side and TEE, such as input buffers, output buffers, and shared memory regions. Specifically, the adversary can: (1) observe and modify outputs sent from TEE to the normal side, (2) craft malicious inputs or parameters before they enter TEE, and (3) alter the contents of shared memory at any time. However, the adversary cannot bypass the hardware-enforced isolation of TEE, nor can they execute arbitrary code inside it without exploiting software vulnerabilities. Under this threat model, we focus on repairing vulnerabilities that stem from bad partitioning between the normal side and TEE, including unencrypted data outputs, insufficient input validation, and unsafe reliance on shared memory.

\subsubsection{Unencrypted Data Output}
A common bad partitioning issue occurs when sensitive data is transferred from TEE to the normal side without encryption. Since the normal-side environment is fully under adversarial control, any plaintext data copied outside TEE can be immediately accessed by the attacker. This results in direct leakage of confidential assets such as cryptographic keys, authentication tokens, or internal state information. Even seemingly non-critical data, when leaked in plaintext—for instance through logs—can enable inference attacks~\cite{10.1145/3707453}, where adversaries analyze data variations to deduce sensitive information. To mitigate this risk, sensitive outputs should always be protected by secure mechanisms, such as encryption, before leaving TEE.

\subsubsection{Input Validation Weaknesses}
Another prevalent issue is insufficient validation of inputs passed from the normal side into TEE. Although TEE SDKs typically enforce basic checks (e.g., preventing null pointers), they lack semantic understanding of how inputs are used inside the secure application. Without explicit validation such as verifying buffer sizes, value ranges, or data formats, TEE code becomes vulnerable to memory safety violations, including buffer overflows and data corruption~\cite{10.1145/3373376.3378486}. Proper input validation must therefore be implemented as a security mechanism to prevent attackers from exploiting malformed or malicious inputs.

\subsubsection{Direct Usage of Shared Memory}
Shared memory is a region of physical memory that can be simultaneously accessed by both the normal side and TEE to exchange data efficiently. It provides a zero-copy mechanism that avoids redundant copying of buffers across trust boundaries, making it a widely used communication channel in TEE applications. However, its direct use can compromise integrity. Since the normal side can arbitrarily modify shared memory contents at runtime, TEE code that directly reads or processes shared memory without integrity verification is exposed to tampering attacks~\cite{10.1145/3407023.3407072, 9309302, DBLP:conf/ndss/ChenLMLC024}. To ensure correctness and security, developers should validate shared memory contents before use or adopt secure copying strategies that decouple TEE logic from untrusted memory regions.

\begin{table}[t]
\centering
\caption{Metadata of projects in \ccBenchName.}
\label{tab:benchmark}
\scriptsize
\setlength{\tabcolsep}{1.2mm}
\begin{tabular}{lrrrrrp{5.5cm}}
\toprule
\textbf{Project} & \textbf{LoC} & \textbf{Stars} & \textbf{Commits} & \textbf{Contributors} & \textbf{Issues} & \textbf{Description} \\
\midrule
luckfox-pico    & 178,627 & 611 & 75   & 6   & 277 & Real-world embedded SDK supporting TEE \\
PPFL            & 80,589 & 74  & 11   & 3   & 4   & Privacy-preserving federated learning with TEE \\
darknetz        & 34,484 & 94  & 180  & 1   & 31  & Deep Neural Network model in TEE \\
optee\_apps     & 10,222 & 13  & 183  & 14  & 2   & TEE applications for secure communication \\
optee\_examples & 5,380 & 201 & 99   & 24  & 57  & Official OP-TEE samples \\
basicAlg\_use   & 5,162 & 8   & 19   & 2   & 0   & Wrapper for TEE encryption algorithm \\
DYMKM           & 1,626 & 147 & 23   & 3   & 0   & An implementation of OP-TEE shared memory \\
Lenet5\_in\_OPTEE & 1,610 & 9 & 32   & 1   & 1   & A Lenet5 application in TEE \\
optee-sdp       & 830 & 3   & 5    & 1   & 0   & Secure device storage \\
\bottomrule
\end{tabular}
\end{table}

\begin{table*}[t]
    \caption{Test Cases for \ccBenchName, where $x$ and $y$ represent the data of output, input and shared memory, $\texttt{len}(x)$ gives the length of $x$, $n$ is the bounds check value, and $x \stackrel{\texttt{NS}}{\rightsquigarrow} y$ indicates that the data $x$ is changed to $y$ in the normal side.}
    \label{tbl:orcale}
    \renewcommand{\arraystretch}{1.3}
    \scriptsize
    \setlength{\tabcolsep}{1.5mm}
    \centering
	\begin{tabular}{lccp{5cm}}
        \toprule
        \makecell[c]{\textbf{Repaired Issues}} & \textbf{Inputs} & \textbf{Expected Outputs} & \makecell[c]{\textbf{Description}}\\
        \midrule
        Unencrypted Data Output & --- & $\texttt{ENC}(x)$ & \makecell[c{p{5cm}}]{The output must be the ciphertext of $x$ obtained by a encryption function $\texttt{ENC}()$} \\
        \midrule
        \multirow{6}{*}{Input Validation Weaknesses}  & $\texttt{len}(x) < n$ & \texttt{TEE\_SUCCESS} & \multirow{3}{5cm}{While an input buffer $x$ will be copied to an in-TEE buffer, the length of $x$ cannot be longer than the in-TEE buffer} \\ 
         & $\texttt{len}(x) = n$ & \texttt{TEE\_SUCCESS} & \\
         & $\texttt{len}(x) > n$ & \texttt{TEE\_ERROR\_BAD\_PARAMETERS} & \\
        \cline{2-4}
         & $\texttt{len}(x) < n$ & \texttt{TEE\_ERROR\_BAD\_PARAMETERS} & \multirow{3}{5cm}{If an input buffer $x$ will be access by an array index, then the length of $x$ should be longer than the array index} \\
         & $\texttt{len}(x) = n$ & \texttt{TEE\_SUCCESS} & \\
         & $\texttt{len}(x) > n$ & \texttt{TEE\_SUCCESS} & \\
        \midrule
        \multirow{2}{*}{Direct Usage of Shared Memory} & $x$ & \texttt{TEE\_SUCCESS} & \multirow{2}{5cm}{It should detect whether the contents of shared memory $x$ have been tampered with to $y$}\\
         & $x \stackrel{\texttt{NS}}{\rightsquigarrow} y$ & \texttt{TEE\_ERROR\_BAD\_PARAMETERS} & \\
        \bottomrule
	\end{tabular}
\end{table*}

\subsection{Benchmark and Test Cases}\label{sec:bgbench}

\textbf{Benchmark Construction.} In this paper, we evaluate \ccSysName using \ccBenchName~\cite{ma2025ditingstaticanalyzeridentifying}, a benchmark for TEE bad partitioning vulnerabilities.
The benchmark construction follows a two-phase approach. First, initial vulnerability cases were identified through manual review of real-world TEE projects collected from GitHub. Second, the dataset was systematically expanded using the CryptoAPI-Bench methodology~\cite{8901573}, which generates comprehensive test cases by introducing variations of vulnerable patterns (such as interprocedural propagation, complex data structures, path-sensitive scenarios, and combined complexity cases) to ensure coverage of diverse vulnerability manifestations.

We acknowledge that the benchmark includes projects of varying maturity levels. To characterize project quality, we adopt the criteria proposed by Munaiah et al.~\cite{10.1007/s10664-017-9512-6} for identifying engineered software projects, which classifies repositories based on dimensions such as LoC, community engagement, documentation quality, commit history, and issue tracking. Table~\ref{tab:benchmark} summarizes the metadata of projects in the benchmark. The benchmark includes both well-maintained projects and smaller experimental ones to reflect the realistic spectrum of TEE application development. This diversity provides both ecological validity from real-world bugs and comprehensive coverage from systematically generated cases, enabling us to evaluate the effectiveness of \ccSysName across varying code quality levels. In addition, the original \ccBenchName contains 110 cases with both vulnerable and non-vulnerable code. To ensure dataset validity, two TEE security experts independently labeled all cases. Expert A identified 85 vulnerable cases, while Expert B identified 90. The two experts agreed on 85 cases as vulnerable and 20 cases as non-vulnerable, with a Cohen’s Kappa of 0.86. After discussion with a third expert to resolve disagreements, we finalized 89 validated vulnerable cases for repair evaluation.

\noindent\textbf{Test Case Construction.} Since the original projects do not provide test cases, we constructed test cases for \ccBenchName through a two-layer approach, ensuring 100\% vulnerability coverage. For interface reachability, we statically analyze TEE source code to extract command IDs and parameter specifications from entry-point handlers, automatically generating test clients that invoke all valid TEE interfaces. This ensures that every vulnerable function entry point is reachable without manual intervention, as described in Section~\ref{sec:tcases}. For vulnerability-specific validation, we manually constructed 172 test cases (covering all 89 vulnerabilities) with inputs specifically crafted to trigger vulnerable functions and expected outputs that verify security properties after repair, as illustrated in Table~\ref{tbl:orcale}: (1) encryption test cases verify correct ciphertext outputs, (2) input validation test cases include both valid and invalid boundary values, and (3) shared memory test cases verify deep copy and integrity check patterns. These test cases were cross-validated by two authors to ensure correctness.

\section{Methodology} \label{sec:method}

\begin{figure*}[t]
    \centering
    \includegraphics[width=0.85\linewidth]{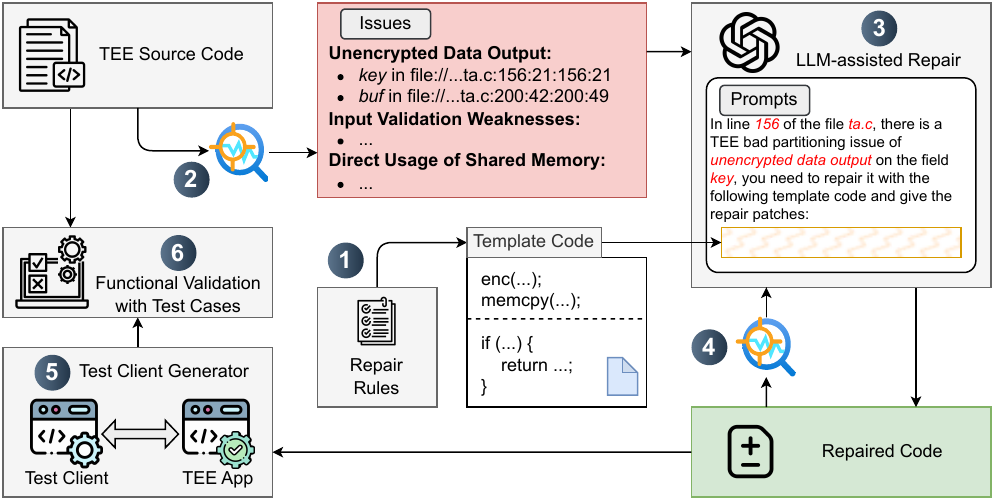}
    \caption{Overview workflow of \ccSysName. The 6 steps are: \blackding{1} template code generating; \blackding{2} issue locating; \blackding{3} LLM-assisted repair; \blackding{4} feedback; \blackding{5} test client generating; \blackding{6} functional validation.}
    \label{fig:overview}
\end{figure*}

As shown in Fig.~\ref{fig:overview}, \ccSysName consists of 6 steps:
\blackding{1} We provide three rules for the repair solutions of TEE bad partitioning issues. Based on these rules, we give the template code that will be used in the following steps.
\blackding{2} Using the bad partitioning detection tool for the TEE application to get the issue report.
\blackding{3} By injecting the related information into LLM prompts and submitting the TEE source code and the corresponding repair template according to vulnerability type and code, the LLM is required to generate a repair patch.
\blackding{4} Using the bad partitioning detection tool to analyze the repaired code again to ensure that the issues are fixed, and feed the detection results back to the LLM.
\blackding{5} In order to validate the functional correctness of the repair results, we can automatically build a test client based on the repaired source code to execute test cases.
\blackding{6} We need to validate whether the output of the repaired TEE application meets the expected output of the test case to show that the repaired code does not change the original functionality of the TEE application.

As introduced in Section~\ref{sec:definition}, prior studies have identified three types of bad partitioning issues in TEEs: unencrypted data outputs, input validation weaknesses, and direct usage of shared memory. Ma et al.~\cite{ma2025ditingstaticanalyzeridentifying} proposed a static analysis tool that systematically detects these issues in TEE projects. In this work, we integrate their detection tool into our repair pipeline. However, detection alone is insufficient to achieve repairs, as it only locates vulnerable code fragments without providing guidance on how to fix them. To bridge this gap, our key idea is to design a domain-specific language (DSL) that allows developers and debuggers to express repair rules as structured transformations. Each DSL rule encodes insecure partitioning patterns and their secure alternatives, and can be automatically instantiated into repair templates. These templates are then combined with large language models (LLMs), which infer context-specific details such as variable names, buffer sizes, or cryptographic parameters, thereby generating semantically correct patches. 
The following section presents our DSL design, demonstrates how it instantiates repair templates for the three types of bad partitioning issues (Section~\ref{sec:DSL}), explains how it guides automated repair using LLMs (Section~\ref{sec:llms}), and introduces the two-stage validation pipeline for repair patches (Section~\ref{sec:tcases}). This design systematically addresses the three challenges identified in Section 1:
\begin{itemize}[leftmargin=*]
    \item \textbf{Challenge 1 (Lack of Guidelines):} Our DSL encodes expert knowledge about secure TEE partitioning practices into structured repair templates, serving as reliable ground truths for patch generation.
    \item \textbf{Challenge 2 (Missing Semantics):} The LLM resolves context-dependent semantic information by interpreting local data-flow and natural-language cues, which traditional static analysis cannot reliably obtain.
    \item \textbf{Challenge 3 (No Feedback Loop):} A two-stage strategy decouples repair from hardware constraints: static analysis provides fast feedback during repair, while generated test clients validate patches on real hardware.
\end{itemize}

\subsection{DSL-Based Repair Templates}
\label{sec:DSL}

\begin{figure}[t]
    \centering
    \begin{lstlisting}[language=DSL]
Program      ::= Node | Node ";" Program
Node         ::= FuncCall | Guard
FuncCall     ::= COPY "(" Dest "," Src "," Len ")"
               | SNPRINT "(" Dest "," Format "," Args ")"
               | MALLOC "(" Size ")" "→" Base
               | ENC "(" Plain "," Cipher "," Len ")"
               | MULMALLOC"(" Args ")" "→" Bases
               | MULENC"(" Args "," Ciphers ")"
               | ARRAY "(" Base "," Index ")" "→" Var
               | SHALLOW "(" Base ")" "→" Var
               | READ "()" "→" Var
               | WRITE "(" Var ")"
               | HASH "(" Base "," Var "," Len ")"
               | MUTATE "(" Var ")" "→" Base
Guard        ::= IF "(" LeftVar "," RelOp "," RightVar ")" "→" Action

Var/Len/Format/Size/Index/LeftVar/RightVar    ::= identifiers | literals
Base(s)/Dest/Src/Plain/Cipher(s)              ::= identifiers.address

Args         ::= Arg | Arg "," Args
Arg          ::= Var
RelOp        ::= "==" | "!=" | "<" | "<=" | ">" | ">="
Action       ::= return ECODE

MULMALLOC    ::= MALLOC | MALLOC ";" MULMALLOC
MULENC       ::= ENC | ENC ";" MULENC
\end{lstlisting}
    \caption{Domain-Specific Language Grammar.}
    \label{fig:dsl}
\end{figure}

The DSL addresses \textbf{Challenge 1} by encoding established secure TEE partitioning practices into structured templates. These patterns are derived from security guidelines~\cite{sgx-guide, optee-guide, iTrustee, optee} and real-world vulnerability analyses~\cite{ma2025ditingstaticanalyzeridentifying, niu2026trustinsecuredemystifyingdevelopers}, providing reliable ground truths. Fig.~\ref{fig:dsl} defines the grammar of our DSL. A \textbf{Program} is a sequence of \textbf{Node} elements separated by semicolons, representing a linearized view of TEE code. Each \textbf{Node} corresponds to either a function call (\textbf{FuncCall}) or a conditional branch (\textbf{Guard}). In the Node descriptions, symbols such as \texttt{Var}, \texttt{Len}, and \texttt{Format} (line 15 of Fig.~\ref{fig:dsl}) typically denote variable names, numeric values, or constant strings. Symbols in line 16 (\eg, \texttt{Base} and \texttt{Dest}) represent aliases of memory buffer base addresses, i.e., they conceptually refer to the start address of a memory region.

A \textbf{FuncCall} abstracts common memory and computation operations in TEE applications. For instance, \texttt{COPY} copies data from the source buffer (\texttt{Src}) to the destination buffer (\texttt{Dest}), optionally with a specified length (\texttt{Len}). Similarly, \texttt{SNPRINT} copies data to the destination buffer but supports variable arguments (\texttt{Args}), which are formatted into a string according to \texttt{Format}. \texttt{MALLOC} allocates a buffer (\texttt{Base}) of the given \texttt{Size} inside TEE, while \texttt{ENC} performs symmetric encryption of a plaintext buffer (\texttt{Plain}) into a ciphertext buffer (\texttt{Cipher}). For multiple arguments, \texttt{MULMALLOC} and \texttt{MULENC} apply \texttt{MALLOC} and \texttt{ENC} to each argument (\texttt{Arg}) sequentially.

The grammar also models array and memory operations. \texttt{ARRAY(Base, Index)$\rightarrow$Var} denotes accessing element \texttt{Var} at a specific \texttt{Index} of array \texttt{Base}, while \texttt{SHALLOW(Base)$\rightarrow$Var} creates a shallow copy of a memory space. Secure storage interactions are captured by \texttt{READ()$\rightarrow$Var} and \texttt{WRITE(Var)}, representing reading from and writing to secure objects, which are TEE-protected areas for persistent data such as keys or hash values. Additionally, \texttt{HASH(Base, Var, Len)} computes a hash value \texttt{Var} over the input buffer (\texttt{Base}), optionally constrained by length (\texttt{Len}). Finally, \texttt{MUTATE(Var)$\rightarrow$Base} models in-place transformations, where the value of \texttt{Var} is modified and stored back to \texttt{Base}.

A \textbf{Guard} models conditional execution with \texttt{IF(LeftVar, RelOp, RightVar)$\rightarrow$Action}, where a relational operator (\texttt{RelOp}) compares two variables (\texttt{LeftVar} and \texttt{RightVar}) to determine if the guarded \texttt{Action} should be executed. \texttt{RelOp} may be any standard comparison operator, including equality, inequality, and relational comparisons (\eg, \texttt{<}, \texttt{<=}). The \texttt{Action} is expressed as \texttt{return ECODE}, where \texttt{ECODE} denotes an error or status code returned when the guard condition holds.

Overall, this grammar captures the essential semantics of TEE function calls and conditional control flow while abstracting away low-level syntax. It provides a structured and analyzable representation that enables us to systematically encode insecure programming patterns and their secure counterparts. In particular, the DSL serves as the foundation for defining repair rules, instantiating them into patch templates, and guiding the LLM to generate context-aware fixes that are both syntactically valid and semantically correct.

With the above grammar, we can formally encode bad partitioning patterns and their corresponding secure transformations. Figs.~\ref{fig:rule1} to~\ref{fig:rule3} illustrate three representative repair rules, each consisting of two parts: a DSL-described rule and the corresponding patch template generated from it. A rule is divided into two segments by the symbol $=>$. The upper part denotes the vulnerable code pattern identified by the detection tool, which we call the trigger. The lower part specifies the secure replacement, referred to as the transformer.
With the defined rules, our DSL will run to generate concrete patch templates by aligning the repair transformations with the actual source code. In these templates, deleted lines are prefixed with $-$, while newly added lines are prefixed with $+$. The original lines affected by the bad partitioning issue are highlighted, corresponding to the triggers defined in the DSL. By matching fields in the vulnerable code with those in the DSL trigger, most parts of the repaired TEE code can be automatically instantiated from the transformer. For fields that cannot be directly inferred (e.g., buffer sizes or variable names), we retain their symbolic representation in the DSL using a placeholder prefix (\$), which will subsequently be resolved by the LLM.

\begin{figure*}[t]
  \centering
  \begin{subfigure}[c]{0.42\linewidth}
    \centering
    \caption{Rule 1.1 (DSL)}\label{fig:dsl1.1}
\begin{lstlisting}[language=DSL, linebackgroundcolor={\ifnum\value{lstnumber}=1\color{cyan!20}\fi}]
COPY($out, $plain, $len) → _
=>
MALLOC($len) -> $cipher;
ENC($plain, $cipher, $len) → _;
COPY($out, $cipher, $len) → _
\end{lstlisting}
  \end{subfigure}\hfill
  \begin{subfigure}[c]{0.53\linewidth}
    \centering
    \caption{Rule 1.1 (Patch Template)}\label{fig:code1.1}
\begin{lstlisting}[language=TEE, linebackgroundcolor={\ifnum\value{lstnumber}=3\color{cyan!20}\fi}]
+ char $cipher[128] = {0};
+ enc(plain, $cipher, 128);
- TEE_MemMove(params[0].memref.buffer, plain, 128);
+ TEE_MemMove(params[0].memref.buffer, $cipher, 128);
\end{lstlisting}
  \end{subfigure}

  \begin{subfigure}[c]{0.42\linewidth}
    \centering
    \caption{Rule 1.2 (DSL)}\label{fig:dsl1.2}
\begin{lstlisting}[language=DSL, linebackgroundcolor={\ifnum\value{lstnumber}=1\color{cyan!20}\fi}]
SNPRINT($out, $format, $args) → _
=>
MULMALLOC($args) → $ciphers;
MULENC($args, $ciphers) → _;
SNPRINT($out, $format, $ciphers) → _
\end{lstlisting}
  \end{subfigure}\hfill
  \begin{subfigure}[c]{0.53\linewidth}
    \centering
    \caption{Rule 1.2 (Patch Template)}\label{fig:code1.2}
\begin{lstlisting}[language=TEE, linebackgroundcolor={\ifnum\value{lstnumber}=5\color{cyan!20}\fi
\ifnum\value{lstnumber}=6\color{cyan!20}\fi}]
+ char $cipher0[strlen(arg0)] = {0};
+ char $cipher1[strlen(arg1)] = {0};
+ enc(arg0, $cipher0, strlen(arg0));
+ enc(arg1, $cipher1, strlen(arg1));
- snprintf(params[0].memref.buffer, params[0].memref.
      size, "%s-%s", arg0, arg1);
+ snprintf(params[0].memref.buffer, params[0].memref.
      size, "%s-%s", $cipher0, $cipher1);
\end{lstlisting}
  \end{subfigure}
  \caption{Rule 1: Unencrypted data output repaired by adding encryption.}
  \label{fig:rule1}
\end{figure*}

\begin{figure*}[t]
  \centering
  \begin{subfigure}[c]{0.45\linewidth}
    \centering
    \caption{Rule 2.1 (DSL)}\label{fig:dsl2.1}
\begin{lstlisting}[language=DSL, linebackgroundcolor={\ifnum\value{lstnumber}=1\color{cyan!20}\fi}]
COPY($dst, $in, $len) → _
=>
IF($len, >, $value) → return ECODE;
COPY($dst, $in, $len) → _
\end{lstlisting}
  \end{subfigure}\hfill
  \begin{subfigure}[c]{0.5\linewidth}
    \centering
    \caption{Rule 2.1 (Patch Template)}\label{fig:code2.1}
\begin{lstlisting}[language=TEE, linebackgroundcolor={\ifnum\value{lstnumber}=4\color{cyan!20}\fi
\ifnum\value{lstnumber}=5\color{cyan!20}\fi}]
+ if (params[1].memref.size > $value) {
+     return TEE_ERROR_BAD_PARAMETERS;
+ }
  TEE_MemMove(buf, params[1].memref.buffer, 
      params[1].memref.size);
\end{lstlisting}
  \end{subfigure}
  \begin{subfigure}[c]{0.45\linewidth}
    \centering
    \caption{Rule 2.2 (DSL)}\label{fig:dsl2.2}
\begin{lstlisting}[language=DSL, linebackgroundcolor={\ifnum\value{lstnumber}=1\color{cyan!20}\fi}]
ARRAY($base, $index) → _
=>
IF($index, >, $value) → return ECODE;
IF($index, <, 0) → return ECODE;
ARRAY($base, $index) → _
\end{lstlisting}
  \end{subfigure}\hfill
  \begin{subfigure}[c]{0.5\linewidth}
    \centering
    \caption{Rule 2.2 (Patch Template)}\label{fig:code2.2}
\begin{lstlisting}[language=TEE, linebackgroundcolor={\ifnum\value{lstnumber}=7\color{cyan!20}\fi}]
+ if (params[0].value.a - 8 > $value) {
+     return TEE_ERROR_BAD_PARAMETERS;
+ }
+ if (params[0].value.a - 8 < 0) {
+     return TEE_ERROR_BAD_PARAMETERS;
+ }
  array[params[0].value.a - 8] = 43;
\end{lstlisting}
  \end{subfigure}
  \caption{Rule 2: Input validation weaknesses repaired by inserting a guard.}
  \label{fig:rule2}
\end{figure*}

\begin{figure*}[t]
  \centering
  \begin{subfigure}[c]{0.45\linewidth}
    \centering
    \caption{Rule 3.1 (DSL)}\label{fig:dsl3.1}
\begin{lstlisting}[language=DSL, linebackgroundcolor={\ifnum\value{lstnumber}=1\color{cyan!20}\fi}]
SHALLOW($sm) → $buf
=>
MALLOC($size) → $buf;
COPY($buf, $sm, $size) → _;
READ() → $h1;
HASH($buf, $h2, $size) → _;
IF(equal($h1,$h2), !=, 0) → return ECODE
\end{lstlisting}
  \end{subfigure}\hfill
  \begin{subfigure}[c]{0.5\linewidth}
    \centering
    \caption{Rule 3.1 (Patch Template)}\label{fig:code3.1}
\begin{lstlisting}[language=TEE, linebackgroundcolor={\ifnum\value{lstnumber}=1\color{cyan!20}\fi}]
- char* buf = params[3].memref.buffer;
+ char buf[$size] = {0};
+ TEE_MemMove(buf, params[3].memref.buffer, $size);
+ char $hash1[256];
+ read($hash1);
+ char $hash2[256];
+ hash($hash2, buf, $size);
+ if (TEE_MemCompare($hash1, $hash2, 256) != 0) {
+     return TEE_ERROR_BAD_PARAMETERS;
+ }
\end{lstlisting}
  \end{subfigure}
\begin{subfigure}[c]{0.45\linewidth}
    \centering
    \caption{Rule 3.2 (DSL)}\label{fig:dsl3.2}
\begin{lstlisting}[language=DSL, linebackgroundcolor={\ifnum\value{lstnumber}=1\color{cyan!20}\fi}]
MUTATE($value) → $sm
=>
MUTATE($value) → buf;
HASH($buf, $h, $size) → _;
WRITE($h$) → _;
COPY($sm, $buf, $size) → _
\end{lstlisting}
  \end{subfigure}\hfill
  \begin{subfigure}[c]{0.5\linewidth}
    \centering
    \caption{Rule 3.2 (Patch Template)}\label{fig:code3.2}
\begin{lstlisting}[language=TEE, linebackgroundcolor={\ifnum\value{lstnumber}=1\color{cyan!20}\fi}]
- *(params[3].memref.buffer + 10) = 0x55;
+ *($buf + 10) = 0x55;
+ char $hash[256];
+ hash($hash, $buf, $size);
+ write($hash);
+ TEE_MemMove(params[3].memref.buffer, $buf, $size);
\end{lstlisting}
  \end{subfigure}
  \caption{Rule 3: Direct usage of shared memory repaired by deep copy and integrity check.}
  \label{fig:rule3}
\end{figure*}

\textbf{Repair Rule 1: Unencrypted data output repaired by adding encryption.}
This rule enforces encryption before sensitive data leaves TEE. Rule 1.1 (Fig.~\ref{fig:dsl1.1}) triggers when sensitive data \texttt{plain} is directly copied to output parameter \texttt{out}. The transformer replaces \texttt{COPY}(\texttt{out}, \texttt{plain}, \texttt{len}) with \texttt{COPY}(\texttt{out}, \texttt{cipher}, \texttt{len}) by: 1) declaring buffer \texttt{cipher} for ciphertext; 2) applying symmetric encryption \texttt{ENC} with matching size; 3) copying \texttt{cipher} to the output.
Fig.~\ref{fig:code1.1} shows the generated template. From line 3 (to be deleted), we extract \texttt{out} (\texttt{params[0].memref.buffer}), \texttt{plain} (\texttt{plain}), and \texttt{len} (128). The template declares a 128-byte buffer \texttt{\$cipher} (line 1), encrypts via \texttt{enc()} (line 2), and copies \texttt{\$cipher} to the output (line 4).
For variadic functions like \texttt{snprintf} that output multiple data, Rule 1.2 (Fig.~\ref{fig:dsl1.2}, lines 3 to 4) generates corresponding ciphertext for each argument. The resulting template (Fig.~\ref{fig:code1.2}) expands \texttt{MULMALLOC} and \texttt{MELENC} into multiple memory initialization and encryption statements.

\textbf{Repair Rule 2: Input validation weaknesses repaired by inserting a guard.}
Unvalidated inputs into the TEE may cause memory corruption vulnerabilities such as buffer overflow. To fix this, we generate validation code placed before input usage.
Rule 2 (Fig.~\ref{fig:rule2}) handles two triggers: copying input parameter \texttt{in} to in-TEE buffer \texttt{dst} (Rule 2.1), and accessing array \texttt{base} with an \texttt{index} (Rule 2.2), where the input can be either the array itself or an access index. The transformer constructs conditional statements comparing sizes or checking index values accordingly. In Rule 2.1, input length must not exceed the in-TEE buffer (line 3 of Fig.~\ref{fig:dsl2.1}); in Rule 2.2, the index must be within valid bounds (lines 3 to 4 of Fig.~\ref{fig:dsl2.2}).
The generated templates are shown in Fig.~\ref{fig:code2.1} and Fig.~\ref{fig:code2.2}. Fig.~\ref{fig:code2.1} (lines 1 to 3) adds a value check for \texttt{params[1].memref.size}, returning \texttt{TEE\_ERROR\_BAD\_PARAMETERS} if invalid. Similarly, Fig.~\ref{fig:code2.2} includes bounds checks to prevent array out-of-bounds access.

\textbf{Repair Rule 3: Direct usage of shared memory repaired by deep copy and integrity check.}
Rule 3 (Fig.~\ref{fig:rule3}) replaces shallow copies of shared memory with deep copies and adds integrity verification. Rule 3.1 triggers on shallow copy statements (line 1 of Fig.~\ref{fig:dsl3.1}) and constructs a buffer \texttt{buf} with matching name and size to deep copy data into the TEE. Following existing work~\cite{10477533, 9925569}, we verify integrity by comparing the original hash \texttt{h1} from a secure object with the recalculated hash \texttt{h2} of the copied data.
If the two are inconsistent, it means that the shared memory data has been modified.
The template in Fig.~\ref{fig:code3.1} performs deep copy (line 3) to replace shallow copies (e.g., line 1), with integrity checks in lines 4 to 10. For operations modifying shared memory (line 1 of Fig.~\ref{fig:code3.2}), Rule 3.2 ensures changes are first performed on the deep copy \texttt{buf}, then copied back after updating the secure object. As shown in Fig.~\ref{fig:code3.2}, operations on \texttt{params[3].memref.buffer} are redirected to \texttt{buf} (lines 1 to 2), and lines 3 to 6 update the integrity value and transfer data back to shared memory.

\subsection{LLM-Based Placeholder Synthesis}
\label{sec:llms}

\begin{figure*}[t]
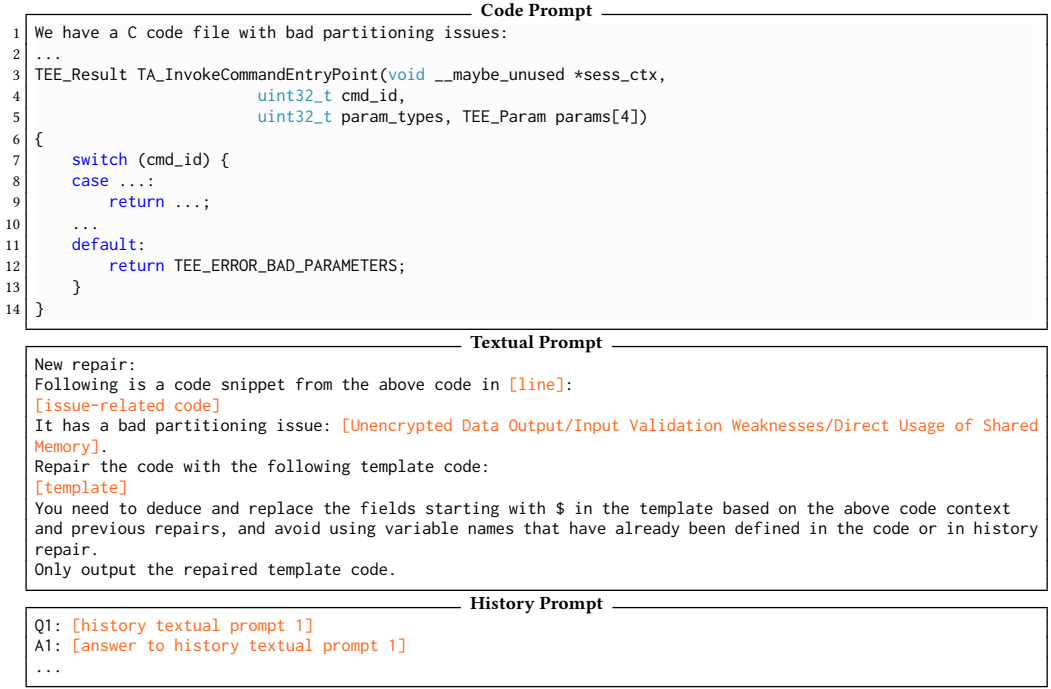

\centering
    \begin{minipage}[t]{\linewidth}
    \begin{CSourceCode*}{xleftmargin=8pt, xrightmargin=3pt,label=\textcolor{black}{\scriptsize \sf \bf Code Prompt}}
We have a C code file with bad partitioning issues:
...
TEE_Result TA_InvokeCommandEntryPoint(void __maybe_unused *sess_ctx,
			uint32_t cmd_id,
			uint32_t param_types, TEE_Param params[4])
{    
    switch (cmd_id) {
    case ...:
        return ...;
    ...
    default:
        return TEE_ERROR_BAD_PARAMETERS;
    }
}
    \end{CSourceCode*}
    \end{minipage}
    \begin{minipage}[t]{\linewidth}\vspace{5pt}
    \begin{MarkdownSourceCode*}{xleftmargin=8pt, xrightmargin=3pt, linenos=false,label=\textcolor{black}{\scriptsize \sf \bf Textual Prompt}}
New repair:
Following is a code snippet from the above code in |\textcolor{R1color}{[line]}|:
|\textcolor{R1color}{[issue-related code]}|
It has a bad partitioning issue: |\textcolor{R1color}{[Unencrypted Data Output/Input Validation Weaknesses/Direct Usage of Shared}|
|\textcolor{R1color}{Memory]}|.
Repair the code with the following template code:
|\textcolor{R1color}{[template]}|
You need to deduce and replace the fields starting with |{\$}| in the template based on the above code context 
and previous repairs, and avoid using variable names that have already been defined in the code or in history 
repair.
Only output the repaired template code.
    \end{MarkdownSourceCode*}
    \end{minipage}
    \begin{minipage}[t]{\linewidth}\vspace{5pt}
    \begin{MarkdownSourceCode*}{xleftmargin=8pt, xrightmargin=3pt, linenos=false,label=\textcolor{black}{\scriptsize \sf \bf History Prompt}}
Q1: |\textcolor{R1color}{[history textual prompt 1]}|
A1: |\textcolor{R1color}{[answer to history textual prompt 1]}|
...
    \end{MarkdownSourceCode*}
    \end{minipage}    
\caption{Prompt for generating the repair patch.}
\label{fig:prompt}
\end{figure*}

Most TEE projects are developed using native languages such as C or C++. However, static analysis tools for C/C++ often struggle with accurately analyzing cases such as dynamically allocated memory sizes and complex pointer behaviors (e.g., \texttt{void *buffer} lacks explicit associations with their semantic constraints, such as corresponding length or index variables). Moreover, due to the strong isolation properties of TEE, applying dynamic analysis to extract runtime stack or heap-related information is also highly constrained or even infeasible. However, the above information is usually needed for repairs, such as patches for buffer size checks or memory copy operations (\textbf{Challenge 2}).
To address these limitations, this section leverages the code understanding capabilities of LLMs to assist in generating repair patches targeted at bad partitioning issues in TEE projects.

As shown in Fig.~\ref{fig:prompt}, we give the prompts to generate repair patches for bad partitioning issues.
The textual prompt includes the line number, issue name, issue-related code, and other information that can help the LLM understand the issue. Then, we also provide the template code defined in Section~\ref{sec:DSL}.
Finally, the LLM is asked to fill in the placeholder fields starting with \$ in the template as the repair patch for the corresponding issue.
In addition, to ensure that LLM-generated code does not have the problem of repeatedly defining variable names after filling the fields of the template, each prompt we upload will also include the complete TEE code and historical questions and answers.
Since the LLM is not required to generate code and only needs to fill in the key parts of the template code, the uncertainty when the LLM repairs the code can be avoided.

\begin{figure*}[t]
    \centering
    \includegraphics[width=0.85\linewidth]{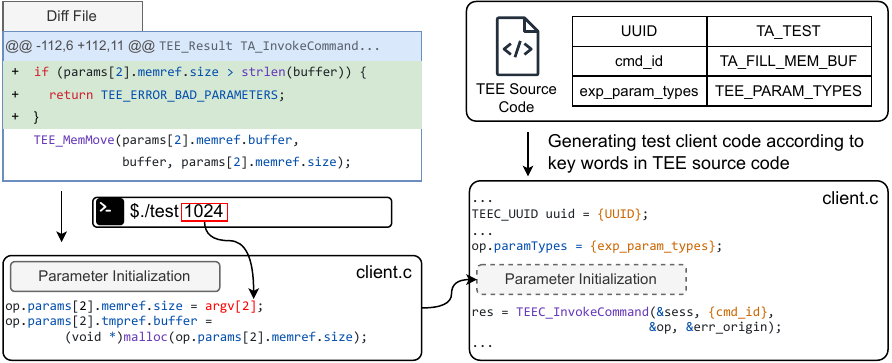}
    \caption{Workflow of generating the test client for a TEE application.}
    \label{fig:gen_client}
\end{figure*}

\subsection{Two-Stage Validation Pipeline}\label{sec:tcases}

Due to the unreliability of LLM-generated results, it is necessary to validate the correctness of any repairs through test cases. These test cases should be used to verify not only whether the repair resolves the TEE bad partitioning issue, but also whether it preserves the original functionality.
However, testing TEE applications presents some challenges. First, each TEE application typically accepts a specific set of parameters, and the accepted parameter types and structures vary across different applications.
Moreover, the normal-side client programs responsible for interacting with the TEE applications are often tightly coupled with the application functions and include large amounts of code unrelated to testing, making them unsuitable for use as test clients. 
In addition, some TEE projects do not even provide client programs.
As a result, there is no client that can be used for testing TEE applications in a general-purpose framework.
What’s more, TEE development requires cross-compilation toolchains and secure hardware deployment, making frequent generate-compile-test iterations quite expensive for each attempt (\textbf{Challenge 3}). \ccSysName adopts a two-stage strategy to decouple repair from hardware constraints:

\textbf{Stage 1: Repair Phase (Static Feedback).} The static analyzer DITING [40] acts as a fast proxy oracle. After the LLM generates a candidate patch, DITING re-analyzes the patched code to check whether vulnerable patterns have been eliminated. If vulnerabilities persist, feedback is provided to the LLM for refinement. This iterative process operates entirely on source code without requiring hardware execution, enabling rapid patch iteration.

\textbf{Stage 2: Validation Phase (Runtime Signals).} Once static analysis confirms the absence of vulnerable patterns, patches are deployed to real TEE hardware for functional validation. \ccSysName automatically generates test clients that invoke the patched TEE Application through standard TEE interfaces. This phase verifies that patches (1) compile without errors under the target TEE SDK, (2) execute successfully on secure hardware, and (3) produce expected outputs that match the intended security semantics (e.g., encrypted data for sensitive outputs, proper validation results for untrusted inputs and shared memory). Any failures observed are leveraged to refine DSL rules for future repairs.

Fig.~\ref{fig:gen_client} shows the workflow for generating test clients and test cases used to validate TEE repairs. The process begins with analyzing the diff file to identify changes introduced in the patched code. For instance, a newly added condition may enforce that the memory size of the input parameter \texttt{params[2].memref} must not exceed the length of a buffer, as shown in the diff segment. Based on this constraint, we need to provide test cases, such as specific values passed via command-line arguments (e.g., 1024) for parameter initialization to trigger the repaired boundary condition.

Meanwhile, metadata (\eg, the application UUID, command ID, and parameter type) is extracted from the TEE source code. This information is used to automatically construct a test client (client.c), which encapsulates only the minimal logic required to invoke the trusted function under test. The client sets up parameters based on user input, allocates memory accordingly, and calls \texttt{TEEC\_InvokeCommand()} with related metadata.

Finally, with the test cases in Section~\ref{sec:bgbench}, we can validate the functional correctness of the repaired code.
For example, for the issues of unencrypted data output, we should confirm whether the TEE application returns the expected output which is correctly encrypted. For input validation weaknesses, the invalid or boundary-triggering inputs (\eg, excessively large memory sizes) must be rejected by the patched code, and the TEE application should return the appropriate error code, such as \texttt{TEE\_ERROR\_BAD\_PARAMETERS}.

\section{Evaluations} \label{sec:eva}
To evaluate the effectiveness of our proposed tool, we have conducted several experiments to answer the following research questions:

\begin{enumerate}[label=\textbf{RQ\arabic*.}, leftmargin=1.3cm]
    \item How accurate is \ccSysName in targeting issue locations compared to the basic LLM?
    \item How effective is \ccSysName at repairing the TEE bad partitioning issues on the benchmark?
    \item How do the individual components of \ccSysName contribute to its repair effectiveness?
    \item How does \ccSysName perform on real-world TEE projects?
\end{enumerate}

\subsection{Environmental Setups}
We implemented the transformation from DSL rules to patch templates in Python. To support the generation of repair patches for TEE applications, we leveraged the \texttt{gpt-4.1-mini-2025-04-14} model. In addition, we combined CodeQL~\cite{githubCodeQL}, a widely used semantic code analysis engine that allows writing queries over program abstract syntax trees and control/data-flow graphs, with Python scripts to extract semantic information from TEE applications and to automatically generate corresponding test clients. All patch generation and test client creation were performed on a server running Ubuntu 24.04, equipped with a 48-core 2.3 GHz AMD EPYC 7643 processor and 512 GB of RAM. The compiled TEE applications and test clients were then deployed and executed on a Raspberry Pi 3B board with a hardware-backed TEE. Finally, the functional correctness of the generated patches was evaluated against a comprehensive suite of test cases.

\noindent\textbf{Baseline Selection.} For comparison, we employed a conversational GPT system with question–answer interactions as a baseline, following the design of the prior LLM-based APR tool, ChatRepair~\cite{xia2023conversationalautomatedprogramrepair,10.1145/3650212.3680323}.
We also compare with representative APR tools, including TBar~\cite{10.1145/3293882.3330577} and Prophet~\cite{10.1145/2837614.2837617} (template-based), and AVR tools like VulMaster\cite{10.1145/3597503.3639222} (learning-based), SAN2PATCH~\cite{DBLP:conf/uss/Kim0KY25} (LLM-based). These tools were selected because their repair processes do not rely entirely on code compilation and execution, thereby avoiding the challenges of TEE cross-compilation and hardware deployment. Since these tools were not designed for TEE, we made adaptations: (1) TBar, Prophet, and VulMaster require precise vulnerability localization, so we provided them with exact line numbers of vulnerable code from DITING~\cite{ma2025ditingstaticanalyzeridentifying}; we also extended TBar and Prophet with DSL-defined templates of \ccSysName while preserving their original patch generation algorithms; for VulMaster, we used its original pre-trained model due to the lack of TEE-specific training data (Challenge 1); (2) ChatRepair relies on its LLMs to infer vulnerable locations; for SAN2PATCH, we replaced sanitizer reports with static analysis reports from DITING~\cite{ma2025ditingstaticanalyzeridentifying} for vulnerability localization, as sanitizers cannot be deployed in TEEs due to hardware isolation. The limitations of existing AVR tools for TEE repair are discussed in Section~\ref{sec:avr}.

\subsection{RQ1: Accuracy of Targeting Issue Locations} \label{sec:rq1}

\begin{table}[t]
    \caption{The precision and recall of issue locating by \ccSysName and the LLM on \ccBenchName. NI is the number of partitioning issues in the benchmark, N is the number of results that tools report, and TP is the number of true positives. P, R and F1 indicates the precision, recall, and F1 score, respectively*.}
    \label{tbl:recall_res}
    \scriptsize
    \centering
    \begin{threeparttable}
        \begin{tabular}{cc|ccccc|ccccc}
        \toprule
        \multirow{2}{*}{\textbf{Bad Partitioning Issues}} & \multirow{2}{*}{\textbf{NI}} & \multicolumn{5}{c|}{\textbf{\ccSysName}} & \multicolumn{5}{c}{\textbf{LLM-based Localization}}\\
         & & \textbf{N} & \textbf{TP} & \textbf{P(\%)} & \textbf{R(\%)} & \textbf{F1} & \textbf{N} & \textbf{TP} & \textbf{P(\%)} & \textbf{R(\%)} & \textbf{F1}\\
        \midrule
        Unencrypted Data Output & 35 & 33 & 33 & 100 & 94.29 & 0.97 & 35 & 35 & 100 & 100 & 1 \\
        Input Validation Weaknesses & 29 & 23 & 23 & 100 & 79.31 & 0.88 & 19 & 19 & 100 & 66.52 & 0.79 \\
        Direct Usage of Shared Memory & 25 & 29 & 25 & 86.21 & 100 & 0.93 & 5 & 0 & 0 & 0 & 0 \\
        \midrule
        \makecell[c]{\textbf{Total}} & 89 & 85 & 81 & 95.29 & 91.01 & 0.93 & 59 & 54 & 91.53 & 60.67 & 0.73\\
        \bottomrule
        \end{tabular}
    \begin{tablenotes}
        \item[*] Note: $P(\%) = TP / N \times 100$, $R(\%) = TP / NI \times 100$.
    \end{tablenotes}
    \end{threeparttable}
\end{table}

Since localization techniques remain inadequate in the existing tools~\cite{DBLP:conf/uss/Hu0SGZXY025}, we evaluate \ccSysName and a LLM-based vulnerability localization method on \ccBenchName, which contains 89 bad partitioning issues. 
As shown in Table~\ref{tbl:recall_res}, \ccSysName achieves superior performance with 95.29\% precision and 91.01\% recall (F1 = 0.93), significantly outperforming the LLM (91.53\% precision, 60.67\% recall, F1 = 0.73). The only exception is unencrypted data outputs, where the LLM attains perfect precision and recall. However, this advantage is limited to a narrow parameter-passing pattern: as shown in Fig.~\ref{fig:copy_code}, when parameters are passed as raw pointers (\texttt{copy}), static analysis loses contextual information, whereas the reasoning ability helps the LLM in this specific case.

\begin{figure}[t]
\centering
\begin{minipage}[c]{0.48\textwidth}
\centering
\begin{CSourceCode*}{firstnumber=1}
void |\color{purple}{copy}|(char *str, int size)
{
    |\color{purple}{TEE\_MemMove}|(str, data, strlen(data));
}
...  
    |\color{purple}{copy}|(params[0].memref.buffer, params[0].memref.size);
\end{CSourceCode*}
\end{minipage}
\hspace{0.02\textwidth}
\begin{minipage}[c]{0.48\textwidth}
\centering
\begin{CSourceCode*}{firstnumber=1}
int a[15] = {...};
int str[15] = {...};

int v = a[params[1].value.a];

char *array = (char *)params[2].memref.buffer;
char c = array[15];
\end{CSourceCode*}
\end{minipage}

\begin{minipage}[c]{0.48\textwidth}
\centering
\caption{Undetected issues of unencrypted data output with \ccSysName.}
\label{fig:copy_code}
\end{minipage}
\hspace{0.02\textwidth}
\begin{minipage}[c]{0.48\textwidth}
\centering
\caption{Undetected issues of input validation weaknesses with the LLM.}
\label{fig:array_code}
\end{minipage}
\end{figure}

For input validation weaknesses, \ccSysName achieves 100\% precision with 79.31\% recall, matching or exceeding the LLM (100\% and 66.52\%). For direct usage of shared memory, \ccSysName detects all cases (recall = 100\%, F1 = 0.93), while the LLM fails entirely (recall = 0\%).
The LLM misses cases for two reasons. First, it only considers inputs serving as array indexes (line 4 of Fig.~\ref{fig:array_code}), overlooking cases where the input itself is the accessed array (line 7). In the latter case, the buffer size parameter \texttt{params[2].memref.size} does not appear explicitly, causing detection failures. \ccSysName examines all array access indexes and captures these cases. Second, LLMs lack domain knowledge about TEE shared memory, which is declared in normal-world code rather than inside TEE. Five issues reported by the LLM correspond to none of the actual shared memory accesses, whereas \ccSysName correctly flags all such statements.

\ans{\textbf{Answer to RQ1:} \ccSysName is significantly more accurate at targeting issue locations than the LLM. \ccSysName achieves an overall precision of 95.29\% and recall of 91.01\% (F1 = 0.93), whereas the LLM reaches only 91.53\% precision and 60.67\% recall (F1 = 0.73).}

\subsection{RQ2: Correctness of Repair Patches}
To evaluate repair correctness, we run the test cases from Section~\ref{sec:bgbench} on \ccBenchName using the test client generated in Section~\ref{sec:tcases}. A patch is considered plausible if it compiles and passes all tests. We then manually verify whether each plausible patch addresses the root cause of the vulnerability according to the security semantics defined in our DSL rules. Table~\ref{tbl:repair_res} reports the results on 89 issues.
Notably, plausible and correct patch counts are closely aligned across tools. None of the evaluated tools use test feedback to guide patch generation; instead, each relies solely on its own strategy (DSL rules, templates, or learned patterns). Test cases serve to verify whether patches faithfully execute the required security operations (\eg, specific ciphertext outputs, precise boundary conditions, copy-then-use for shared memory). Since these behaviors are deterministic, passing all tests generally indicates correct security semantics.

For Unencrypted Data Output, \ccSysName correctly repairs 33/35 issues (94.3\%), significantly outperforming all baselines. Although we extended TBar and Prophet with DSL-defined encryption templates, they rarely associate TEE vulnerable code with the correct vulnerability type (only 6 successes). VulMaster, lacking TEE-specific training data, fails to generate any correct patches, confirming the limitation of learning-based approaches on emerging vulnerabilities (Challenge 1). SAN2PATCH, designed for memory corruption, relies on crash reports for localization, but TEE encryption vulnerabilities do not crash, leaving SAN2PATCH unable to understand or localize these security issues. ChatRepair produces 11 plausible patches, yet none are semantically correct. As shown in Fig.~\ref{fig:incorrect_code}, it misuses a non-standard in-place encryption function (lines 1-6) from the benchmark codebase, so the second call (line 11) re-encrypts the ciphertext of the first call (line 8), which then cannot be decrypted. The template code of \ccSysName avoids this by allocating fresh memory for each ciphertext.

For Input Validation Weaknesses, \ccSysName produces 20 correct patches (67.0\%), outperforming ChatRepair (12, 41.4\%), TBar (11, 37.9\%), Prophet (6, 20.7\%), SAN2PATCH (5, 17.2\%), and VulMaster (1, 0.03\%). TBar and Prophet achieve better template selection here since boundary checking is a common pattern, but struggle with instantiation due to language-specific differences. TBar, originally for Java, cannot handle C/C++ buffer semantics. It cannot use \texttt{sizeof()} or other pointer arithmetic to determine buffer lengths, which Java achieves with the \texttt{length()} method. For example, when repairing a buffer overflow vulnerability with a dynamically allocated buffer declared as \texttt{void *buffer}, it generates \texttt{if (len <= sizeof(buffer))}, which only obtains the pointer size (typically 4 or 8 bytes) rather than the actual allocated buffer size. Prophet targets C/C++ and handles static arrays but also fails on dynamic buffers. \ccSysName instead leverages LLM-based placeholder synthesis to infer such semantics accurately. ChatRepair and SAN2PATCH can generate correct validation conditions but lack precise localization without guidance from DITING.

For Direct Usage of Shared Memory, \ccSysName is the only tool that successfully generates correct patches. All baselines fail completely in this category because it requires understanding the semantics of shared memory. \ccSysName leverages DITING to precisely localize code locations where shared memory is misused, then applies a template with LLM-based synthesis to generate correct patches. For example, most baselines insert \texttt{if (shared\_mem != NULL)} before the vulnerable code, treating it as a null-pointer issue, whereas \ccSysName generates \texttt{memcpy(in\_TEE\_buf, shared\_mem, len)} followed by operations on \texttt{in\_TEE\_buf}, correctly enforcing the TEE security model for shared memory.

\begin{table}[t]
    \caption{Patch generation results* of \ccSysName and baselines on \ccBenchName.}
    \label{tbl:repair_res}
    \scriptsize
    \setlength{\tabcolsep}{1mm}
    \centering
    \begin{threeparttable}
        \begin{tabular}{cc
        >{\centering\arraybackslash}p{0.7cm}@{/}>{\centering\arraybackslash}p{0.7cm}@{/}>{\centering\arraybackslash}p{0.7cm}
        >{\centering\arraybackslash}p{0.7cm}@{/}>{\centering\arraybackslash}p{0.7cm}@{/}>{\centering\arraybackslash}p{0.7cm}
        >{\centering\arraybackslash}p{0.7cm}@{/}>{\centering\arraybackslash}p{0.7cm}@{/}>{\centering\arraybackslash}p{0.7cm}
        >{\centering\arraybackslash}p{0.7cm}@{/}>{\centering\arraybackslash}p{0.7cm}@{/}>{\centering\arraybackslash}p{0.7cm}}
        \toprule
        \textbf{Tools} & \textbf{Category} & \multicolumn{3}{c}{\makecell[c]{\textbf{Unencrypted Data} \\ \textbf{Output}}} & \multicolumn{3}{c}{\makecell[c]{\textbf{Input Validation} \\ \textbf{Weaknesses}}} & \multicolumn{3}{c}{\makecell[c]{\textbf{Direct Usage of} \\ \textbf{Shared Memory}}} & \multicolumn{3}{c}{\textbf{Total}} \\
        \midrule
        \multirow{2}{*}{\ccSysName} & \multirow{2}{*}{DSL + LLM} 
        & 33 & 33 (94.3\%) & 33 (94.3\%) & 23 & 20 (67.0\%) & 20 (67.0\%) & 25 & 25 (100\%) & 25 (100\%) & 81 & 78 (87.6\%) & 78 (87.6\%)  \\\hline
        \multirow{2}{*}{TBar}       & \multirow{2}{*}{Template-based} 
        & 33 & 6 (17.1\%)  & 6 (17.1\%)  & 23 & 11 (37.9\%) & 11 (37.9\%) & 25 & 0 (0.0\%)  & 0 (0.0\%)  & 81 & 17 (19.1\%) & 17 (19.1\%) \\\hline
        \multirow{2}{*}{Prophet}    & \multirow{2}{*}{Template-based} 
        & 33 & 0 (0.0\%)   & 0 (0.0\%)   & 23 & 6 (20.7\%)  & 6 (20.7\%)  & 25 & 0 (0.0\%)  & 0 (0.0\%)  & 32 & 6 (0.07\%)  & 6 (0.07\%) \\\hline
        \multirow{2}{*}{VulMaster}  & \multirow{2}{*}{Learning-based} 
        & 33 & 0 (0.0\%)   & 0 (0.0\%)   & 23 & 1 (0.03\%)  & 1 (0.03\%)  & 25 & 0 (0.0\%)  & 0 (0.0\%)  & 81 & 1 (0.01\%)  & 1 (0.01\%) \\\hline
        \multirow{2}{*}{ChatRepair} & \multirow{2}{*}{LLM-based} 
        & 35 & 11 (31.4\%) & 0 (0.0\%)   & 19 & 12 (41.4\%) & 12 (41.4\%) & 0  & 0 (0.0\%)  & 0 (0.0\%)  & 54 & 23 (25.8\%) & 12 (13.4\%) \\\hline
        \multirow{2}{*}{SAN2PATCH}  & \multirow{2}{*}{LLM-based} 
        & 2  & 0 (0.0\%)   & 0 (0.0\%)   & 5  & 5 (17.2\%)  & 5 (17.2\%)  & 3  & 0 (0.0\%)  & 0 (0.0\%)  & 10 & 5  (0.06\%) & 5 (0.06\%) \\
        \bottomrule
        \end{tabular}
    \begin{tablenotes}
        \item Note: \ccBenchName contains 89 issues: 35 unencrypted data output, 29 input validation weaknesses, and 25 direct usage of shared memory.
        \item[*] The numbers are presented in the form of \#Generated patches / \#Plausible patches / \#Correct patches.
    \end{tablenotes}
    \end{threeparttable}
\end{table}

\begin{figure}[t]
\centering
\begin{CSourceCode*}{firstnumber=1,highlightlines={8,11}}
void |\color{purple}{enc}|(char *str, int len)
{
    for (int i = 0; i < len; i++) {
        str[i] = str[i] + 1;
    }
}
...
+    |\color{purple}{enc}|(key, strlen(key));
     |\color{purple}{TEE\_MemMove}|(params[0].memref.buffer, key, strlen(key));

+    |\color{purple}{enc}|(key, strlen(key));
     |\color{purple}{TEE\_MemMove}|(params[1].memref.buffer, key, strlen(key));
\end{CSourceCode*}
\caption{Overfitting repair on the issues of unencrypted data output by ChatRepair.}
\label{fig:incorrect_code}
\end{figure}

\ans{
\textbf{Answer to RQ2:} \ccSysName achieves the highest repair success rate, correctly fixing 78 out of 89 vulnerabilities (87.6\%), significantly outperforming all baselines. TBar, the second-best performer, correctly repairs 17 vulnerabilities (19.1\%), followed by ChatRepair (12, 13.4\%), Prophet (6, 0.07\%), SAN2PATCH (5, 0.06\%), and VulMaster (1, 0.01\%).
}

\subsection{RQ3: The Ablation Study}
To evaluate the contribution of each component, we designed two variants: DSL-only and LLM-only. Table~\ref{tbl:ab_res} shows the results that \ccSysName achieves 78 correct patches (87.6\%), significantly outperforming both DSL-only (20, 22.5\%) and LLM-only (12, 13.4\%). This demonstrates that neither component alone is sufficient, and their combination is essential for effective TEE vulnerability repair.

The DSL-only approach already obtains vulnerability types and locations from DITING, eliminating the template selection challenge faced by TBar and Prophet. However, its primary limitation lies in placeholder synthesis. Without LLM assistance, DSL-only cannot infer which specific variables in TEE code correspond to buffer sizes, pointers, or shared memory regions. This is particularly challenging due to diverse naming conventions and complex parameter structures in the TEE code (e.g., \texttt{params[x].memref.size}).

The LLM-only approach, like ChatRepair, struggles with vulnerability localization. Without guidance from DITING and structured templates, the LLM often fails to identify the correct code locations requiring modification or to use the correct functions to implement the repair. This limitation has been confirmed in our localization accuracy analysis (Section~\ref{sec:rq1}), where the LLM-based approach achieves significantly lower precision compared to DITING-guided methods.

\begin{table}[t]
    \caption{Patch generation results* of the ablation study.}
    \label{tbl:ab_res}
    \scriptsize
    \centering
    \begin{threeparttable}
        \begin{tabular}{c
        >{\centering\arraybackslash}p{0.7cm}@{/}>{\centering\arraybackslash}p{0.7cm}@{/}>{\centering\arraybackslash}p{0.7cm}
        >{\centering\arraybackslash}p{0.7cm}@{/}>{\centering\arraybackslash}p{0.7cm}@{/}>{\centering\arraybackslash}p{0.7cm}
        >{\centering\arraybackslash}p{0.7cm}@{/}>{\centering\arraybackslash}p{0.7cm}@{/}>{\centering\arraybackslash}p{0.7cm}
        >{\centering\arraybackslash}p{0.7cm}@{/}>{\centering\arraybackslash}p{0.7cm}@{/}>{\centering\arraybackslash}p{0.7cm}}
        \toprule
         & \multicolumn{3}{c}{\makecell[c]{\textbf{Unencrypted Data} \\ \textbf{Output}}} & \multicolumn{3}{c}{\makecell[c]{\textbf{Input Validation} \\ \textbf{Weaknesses}}} & \multicolumn{3}{c}{\makecell[c]{\textbf{Direct Usage of} \\ \textbf{Shared Memory}}} & \multicolumn{3}{c}{\textbf{Total}} \\
        \midrule
        \multirow{2}{*}{\ccSysName}
        & 33 & 33 (94.3\%) & 33 (94.3\%) & 23 & 20 (67.0\%) & 20 (67.0\%) & 25 & 25 (100\%) & 25 (100\%) & 81 & 78 (87.6\%) & 78 (87.6\%)  \\
        \multirow{2}{*}{DSL-only}
        & 33 & 9 (25.7\%)  & 9 (25.7\%)  & 23 & 11 (37.9\%) & 11 (37.9\%) & 25 & 0 (0.0\%)  & 0 (0.0\%)  & 81 & 20 (22.5\%) & 20 (22.5\%)  \\
        \multirow{2}{*}{LLM-only}
        & 35 & 11 (31.4\%) & 0 (0.0\%)   & 19 & 12 (41.4\%) & 12 (41.4\%) & 0  & 0 (0.0\%)  & 0 (0.0\%)  & 54 & 23 (25.8\%) & 12 (13.4\%) \\
        \bottomrule
        \end{tabular}
    \begin{tablenotes}
        \item Note: \ccBenchName contains 89 issues: 35 unencrypted data output, 29 input validation weaknesses, and 25 direct usage of shared memory.
        \item[*] The numbers are presented in the form of \#Generated patches / \#Plausible patches / \#Correct patches.
    \end{tablenotes}
    \end{threeparttable}
\end{table}

\ans{
\textbf{Answer to RQ3:} DSL templates encode essential domain knowledge and ensure correct API usage, while LLM synthesis resolves context-dependent placeholders that the DSL-only approach cannot handle.
}

\subsection{RQ4: In-the-Wild Study} \label{sec:itws}
We applied \ccSysName to the TEE projects collected on GitHub and submitted the repairs as 5 pull requests, 2 of which have been confirmed by developers and merged into the main branch.
For example, the project \textit{basicAlg\_use} is an in-TEE library that includes various cryptographic algorithms such as RSA, SHA, AES, and random number generation. 
After detecting an issue of input validation weakness in the project that could lead to a buffer overflow, we used \ccSysName to repair it and successfully generated check conditions for the input data.

Another project is \textit{PPFL}, which is a privacy-preserving federated learning framework with TEE.
The project uses a lot of code to access the input buffer in array form, but does not verify the array index and the size of the input buffer.
After the repair, we generated a patch containing 15 lines of code changes that resolved the issue, and the patch was accepted by the developer.

We also have some issues that we are working on with developers. For instance, the \textit{optee-sdp} project (an application for securely storing device data in the TEE registers) used \texttt{snprintf(dest, size, format, args)} to store log messages in a buffer (\texttt{dest}) without verifying whether the buffer size was sufficient to accommodate the entire log content. In addition, the log outputs were left unencrypted, exposing a large volume of TEE internal register data to the normal side and creating a risk of sensitive information leakage. \ccSysName detected and repaired these issues, ensuring the security of log handling.

\ans{
\textbf{Answer to RQ4:} \ccSysName performs effectively on real-world TEE projects, successfully producing 2 repairs that are accepted and merged by project maintainers.
}

\section{Discussions}\label{sec:tv}
\subsection{LLM and Runtime Cost of \ccSysName}
We measured the number of input and output tokens of the LLM, as well as the execution time to repair each issue in order to evaluate the cost of \ccSysName. On average, the input and output tokens that the LLM used to repair a bad partitioning issue in the \ccBenchName are 16682 and 72, respectively, which correspond to the estimated cost of \$0.007 (input) and \$0.0001 (output) under the \texttt{gpt-4.1-mini-2025-04-14} model. The execution time ranges from approximately 1.01 to 5.62 seconds, and most of the time is spent waiting for the LLM response.

\subsection{Patch Overfitting}
Unlike test-driven repair approaches that risk generating patches satisfying only specific test cases~\cite{DBLP:conf/uss/Hu0SGZXY025}, TEERepair generates patches via DSL-defined security rules derived from established secure TEE partitioning practices. In our workflow, test cases serve solely for post-hoc validation to confirm that the generated patches compile correctly and preserve program functionality, while they do not guide the patch generation process itself. This ensures that TEERepair targets the root cause of bad partitioning issues (e.g., adding encryption for sensitive outputs, inserting validation for untrusted inputs) rather than only satisfying test conditions. Furthermore, our successful evaluation on real-world projects (Section~\ref{sec:itws}) demonstrates that patches generated by TEERepair generalize beyond our benchmark, as several patches have been accepted by upstream maintainers who independently verified their correctness.

\subsection{Security and Privacy Considerations}
TEE security protects data confidentiality and integrity during processing rather than code secrecy. Similar to cryptographic algorithms, public availability of TEE code can enhance security through peer review. For organizations requiring code confidentiality, TEERepair supports local LLM deployment as an alternative to cloud-based APIs.

\subsection{Threats to Validity}
\subsubsection{External Validity}
\ccSysName primarily targets the repair of TEE applications based on ARM TrustZone, and currently lacks evaluation on platforms such as Intel SGX and AMD SEV. However, the DSL defined in this work is platform-agnostic and applicable to any TEE program development language. Therefore, the proposed approach remains broadly generalizable.

\subsubsection{Construct Validity}
Currently, the test cases used in the benchmark for \ccSysName are manually written. Ideally, automated approaches such as fuzzing could be employed to test TEE applications. However, existing techniques~\cite{10179302} struggle to achieve full code coverage in the TEE setting. As a result, manual test case design remains necessary to ensure comprehensive and reliable evaluation. To mitigate potential biases, three experienced developers independently wrote the initial test cases and cross-checked each other’s work. We therefore believe that the test cases used in this work are of high quality, and the threats to construct validity are minimal.

\section{Related Work}\label{sec:rw}
\subsection{Automated Program Repair}
\noindent\textbf{Traditional APR.} Early APR techniques rely on heuristics, constraints, or templates rather than machine learning~\cite{10.1145/3696450}. Genetic programming approaches such as GenProg~\cite{6035728}, Astor~\cite{10.1145/2931037.2948705}, and SimFix~\cite{10.1145/3213846.3213871} explore candidate patches satisfying test suites. Constraint-based methods like Nopol~\cite{10.1109/TSE.2016.2560811}, Cardumen~\cite{DBLP:conf/ssbse/MartinezM18}, and Dynamoth~\cite{10.1145/2896921.2896931} derive patches from program semantics. Template-based tools including TBar~\cite{10.1145/3293882.3330577}, FixMiner~\cite{10.1007/s10664-019-09780-z}, and Avatar~\cite{8667970} transform buggy code using predefined patterns. However, these approaches require substantial manual effort and depend heavily on test cases for validation, which is a critical limitation for TEE applications where test suites are often absent (Section~\ref{sec:intro}).

\noindent\textbf{Learning-Based APR.} Deep learning introduced Neural Machine Translation (NMT)-based approaches~\cite{10.1145/3631974,10.1145/3340544} that treat repair as translation from buggy to correct code. Representative tools include SequenceR~\cite{8827954}, CoCoNut~\cite{10.1145/3395363.3397369}, DLFix~\cite{10.1145/3377811.3380345}, CURE~\cite{9401997}, SelfAPR~\cite{10.1145/3551349.3556926}, RewardRepair~\cite{10.1145/3510003.3510222}, and others~\cite{10.1145/3468264.3468544,10.1145/3510003.3510177,10172781,10.1145/3631974}. While more generalizable than traditional APR, these methods require large-scale training data that is scarce in the TEE domain and risk generating insecure patches in security-critical contexts.

\noindent\textbf{LLM-Based APR.} Recent approaches leverage LLMs pre-trained on massive code corpora~\cite{zhang2024systematicliteraturereviewlarge,10172803,10.1145/3540250.3549101}. Conversation-style tools like ChatRepair~\cite{10.1145/3650212.3680323,xia2023conversationalautomatedprogramrepair} and ContrastRepair~\cite{10.1145/3719345} refine patches through multi-turn interactions, while cloze-style methods such as AlphaRepair~\cite{10.1145/3540250.3549101}, FitRepair~\cite{10298499}, and Repilot~\cite{10.1145/3611643.3616271} fill missing code fragments. We adapt ChatRepair as our baseline due to its strong performance and reproducibility. Agent-based approaches like RepairAgent~\cite{11029914}, AutoCodeRover~\cite{10.1145/3650212.3680384}, and SpecRover~\cite{11029735} show promise~\cite{roychoudhury2025agenticaisoftwarethoughts}, but are still at an early stage for TEE development given the lack of test suites and cross-device deployment overhead.

\subsection{Automated Vulnerability Repair} \label{sec:avr}
Unlike APR, which typically relies on test suites to locate and validate software defects, Automated Vulnerability Repair (AVR) typically leverages sanitizer-detected crashes~\cite{DBLP:conf/uss/Kim0KY25, DBLP:conf/uss/0003G00XM0025} or static analysis~\cite{10.1145/3705310} to localize vulnerabilities. However, due to the hardware isolation of TEEs, sanitizers cannot be deployed to monitor the TEE execution, making crash reports unavailable and crash-based localization infeasible for TEE applications. More fundamentally, TEE partitioning issues such as unencrypted data output and direct usage of shared memory may not cause crashes, but they can violate security properties such as confidentiality and integrity, allowing attackers to steal or tamper with sensitive data without triggering any observable failure. In this work, we rely on the static analyzer DITING~\cite{ma2025ditingstaticanalyzeridentifying} to detect and localize bad partitioning issues in TEE applications. Therefore, this section focuses on analyzing four categories of patch generation methods: search-based, template-based, semantics-based, and learning-based approaches~\cite{DBLP:conf/uss/Hu0SGZXY025}.

\noindent\textbf{Search-based approaches}, such as GenProg~\cite{6035728}, formulate patch generation as a search problem so that code snippets are iteratively modified and recombined into different program variants to produce candidate patches. CrashRepair~\cite{10.1145/3707454} combines crash-guided concolic execution with code mutation search, which can infer repair constraints at specific program locations and search for code mutations satisfying those constraints. However, search-based methods still rely on dynamic feedback for patch evaluation: GenProg requires test suites, while CrashRepair requires crash-inducing inputs. Neither is applicable to TEE bad partitioning issues, which lack comprehensive test suites (Challenge 3) and may not manifest as crashes.

\noindent\textbf{Template-based approaches} leverage pre-defined fix patterns to transform vulnerable code into secure forms. Representative tools include PAR~\cite{6606626} and TBar~\cite{10.1145/3293882.3330577}, which collect fix templates from human-written patches for Java programs. Prophet~\cite{10.1145/2837614.2837617} ranks candidate patch templates for C programs by learning a patch correctness model from successful human patches. TEERepair adopts a similar idea by encoding secure TEE partitioning patterns into DSL-defined repair rules, and we have added TBar and Prophet as baselines in Section 4.3.

\noindent\textbf{Semantics-based approaches} leverage program analysis techniques such as symbolic execution or constraint solving to generate patches with formal guarantees. Dynamic approaches like ExtractFix~\cite{10.1145/3418461} and Concolic program repair~\cite{10.1145/3453483.3454051} use KLEE-based symbolic/concolic execution to extract crash-free constraints from exploit traces. However, KLEE executes LLVM bitcode within a POSIX emulation layer~\cite{klee}, which is incompatible with TEE applications that depend on vendor-specific cross-compilation toolchains (e.g., ARM TrustZone SDKs) and secure hardware for execution. Static approaches like EffFix~\cite{10.1145/3705310} employ Incorrectness Separation Logic (ISL) via the Pulse analyzer to detect and repair memory safety issues, such as null pointer dereferences and double-frees. However, TEE bad partitioning vulnerabilities violate security properties (confidentiality and integrity) rather than memory safety~\cite{ma2025ditingstaticanalyzeridentifying}, placing them outside the scope of ISL-based analysis.

\noindent\textbf{Learning-based approaches} train neural models on historical vulnerability-fix pairs to generate patches. VulRepair~\cite{10.1145/3540250.3549098} employs byte-pair encoding on a CodeT5 pre-trained language model to generate software vulnerability repairs. Similarly, VulMaster~\cite{10.1145/3597503.3639222} utilizes the Fusion-in-Decoder framework with CodeT5, integrating AST structures and CWE expert knowledge to handle code without length limitation. More recently, LLM-based approaches have emerged: SAN2PATCH~\cite{DBLP:conf/uss/Kim0KY25} uses LLMs to comprehend sanitizer logs, localize root causes of vulnerabilities in source code, and generate corresponding patches; PatchAgent~\cite{DBLP:conf/uss/0003G00XM0025} designs an autonomous agent that integrates fault localization, patch generation, and validation, simulating the repair workflow of human experts. However, these approaches face two key limitations in the TEE development. First, they rely on frequent generate-compile-test iterations, which become expensive due to cross-compilation and hardware deployment requirements (Challenge 3). Second, bad partitioning is an emerging vulnerability class that has only recently gained attention from the security community~\cite{ma2025ditingstaticanalyzeridentifying, niu2026trustinsecuredemystifyingdevelopers}, and consequently, no large-scale dataset of such vulnerabilities and their fixes exists for training learning-based models (Challenge 1).

In addition, these AVR systems assume monolithic software with direct memory access, whereas TEE applications span two isolated execution environments and communicate through constrained interfaces such as shared memory and remote procedure calls. These cross-world interactions introduce unique attack surfaces that do not appear in the above AVR settings. A recent SoK paper~\cite{DBLP:conf/uss/Hu0SGZXY025} systematically surveys AVR methods, finding that vulnerability localization accuracy critically impacts repair effectiveness, yet existing localization techniques remain inadequate, and that fuzzing-based and symbolic execution-based tools often produce overfitting patches. TEERepair addresses both: we leverage DITING~\cite{ma2025ditingstaticanalyzeridentifying} for precise localization of TEE bad partitioning vulnerabilities, and generate templates through DSL-defined security rules encoding expert knowledge rather than test-suite guidance, thereby ensuring root-cause fixes. LLMs then synthesize context-aware code to complete these templates, enhancing both their adaptability and scalability. This synergy between rule-based guidance and LLM-assisted synthesis offers a novel perspective on automated repair for security-critical, low-level TEE applications.

\section{Conclusion and Future Work}\label{sec:conc}
In this paper, we presented \ccSysName, a framework for automatically repairing TEE bad partitioning issues. By combining DSL-based secure programming patterns with LLM semantic reasoning, \ccSysName generates context-aware patches that are both syntactically valid and semantically correct. Our evaluation on \ccBenchName demonstrates a repair success rate of 87.6\%, significantly outperforming baselines, and our in-the-wild study confirms practical applicability to real-world projects. 
Future work includes extending DSL with additional security policies and evaluating it on industrial-scale codebases to integrate automated TEE repair into the secure development lifecycle.

\section*{Data Availability}
The replication package is available at: \url{https://github.com/CharlieMCY/TEERepair}.

\begin{acks}
This research is supported by the National Research Foundation, Singapore, and the Cyber Security Agency of Singapore under its National Cybersecurity R\&D Programme (Proposal ID: NCR25-DeSCEmT-SMU). Any opinions, findings and conclusions or recommendations expressed in this material are those of the author(s) and do not reflect the views of the National Research Foundation, Singapore, and the Cyber Security Agency of Singapore.
\end{acks}

\bibliographystyle{ACM-Reference-Format}
\bibliography{sample-base}


\end{document}